\def \SAIT #1 #2 {{\em Mem.\ Soc.\ Astron.\ It.\/} {\bf #1}, #2}
\def \MESS #1 #2 {{\em The Messenger\/} {\bf #1}, #2}
\def \ASTRNACH #1 #2 {{\em Astron. Nach.\/} {\bf #1}, #2}
\def \AAP #1 #2 {{\em Astron. Astrophys.\/} {\bf #1}, #2}
\def \AAL #1 #2 {{\em Astron. Astrophys. Lett.\/} {\bf #1}, L#2}
\def \AAR #1 #2 {{\em Astron. Astrophys. Rev.\/} {\bf #1}, #2}
\def \AAS #1 #2 {{\em Astron. Astrophys. Suppl. Ser.\/} {\bf #1}, #2}
\def \AJ #1 #2 {{\em Astron. J.\/} {\bf #1}, #2}
\def \ANNREV #1 #2 {{\em Ann. Rev. Astron. Astrophys.\/} {\bf #1}, #2}
\def \APJ #1 #2 {{\em Astrophys. J.\/} {\bf #1}, #2}
\def \APJL #1 #2 {{\em Astrophys. J. Lett.\/} {\bf #1}, L#2}
\def \APJS #1 #2 {{\em Astrophys. J. Suppl.\/} {\bf #1}, #2}
\def \APSS #1 #2 {{\em Astrophys. Space Sci.\/} {\bf #1}, #2}
\def \ASR #1 #2 {{\em Adv. Space Res.\/} {\bf #1}, #2}
\def \BAIC #1 #2 {{\em Bull. Astron. Inst. Czechosl.\/} {\bf #1}, #2}
\def \JSQRT #1 #2 {{\em J. Quant. Spectrosc. Radiat. Transfer\/} {\bf #1},
#2}
\def \MN #1 #2 {{\em Mon. Not. R. Astr. Soc.\/} {\bf #1}, #2}
\def \MEM #1 #2 {{\em Mem. R. Astr. Soc.\/} {\bf #1}, #2}
\def \PLR #1 #2 {{\em Phys. Lett. Rev.\/} {\bf #1}, #2}
\def \PASJ #1 #2 {{\em Publ. Astron. Soc. Japan\/} {\bf #1}, #2}
\def \PASP #1 #2 {{\em Publ. Astr. Soc. Pacific\/} {\bf #1}, #2}
\def \NAT #1 #2 {{\em Nature\/} {\bf #1}, #2}

\documentstyle{memsait}
\input epsf.sty
\begin{opening}
\title{FOSSIL DISKS \& PROPELLER SPINDOWN OF SGR/AXPS}
\author{Richard E. Rothschild$^1$, David Marsden$^2$, Richard E.
Lingenfelter$^1$} \institute{$^1$UCSD/Center for Astrophysics and
Space Sciences, La Jolla, CA\\ $^2$NASA/Goddard Space Flight Center,
Greenbelt, MD} \date{} % DO NOT INSERT ANY DATE HERE !!!
\end{opening}

\begin{document}

%\oddpagefooter{\sf Mem. S.A.It., Vol. ??, ??}{}{\thepage}
%\evenpagefooter{\thepage}{}{\sf Mem. S.A.It., Vol. ??, ??}
\oddpagefooter{}{}{} % LEAVE AS IT IS !
\evenpagefooter{}{}{} % LEAVE AS IT IS !
\
\bigskip

\begin{abstract}
We have shown that the interstellar media which surround the
progenitors of SGRs and AXPs were unusually dense compared to the
environments around most young radio pulsars.
This environmental correlation argues strongly against
the current magnetar model for SGRs and AXPs.
We suggest instead that they are neutron stars with sub-critical
magnetic fields and are spun down rapidly by ``propeller'' torques
from fossil disks formed from the fallback of supernova ejecta.
We show that this hypothesis is consistent with the observed properties
of these enigmatic objects, and we compare the propeller and magnetar
models for SGR and AXPs.
\end{abstract}

\section{Introduction}

Two lines of thought exist as to the nature of Soft Gamma-ray Repeaters
(SGRs) and Anomalous X-ray Pulsars (AXPs). On the one hand, Thompson and
Duncan (1995) propose neutron stars with super-critical ($>10^{14}$ Gauss)
magnetic fields, which spindown the stars and power the gamma-ray bursts.
On the other hand, several authors (van Paradijs et al. 1995; Alpar, 2000;
Chatterjee, Hernquist \& Narayan, 2000; Marsden et al. 2001 [MLRH])
propose neutron stars with typical pulsar magnetic fields ($\sim
10^{12}$ Gauss), which are spundown by magnetospheric ``propeller" torques
from fallback or fossil disks.

The association with visible supernova remnants has long been recognized
(e.g. Cline et al. 1982), and recently reaffirmed (MLRH; Figure 1 and Table I), as the primary
evidence for the SGR/AXP's relatively young ages, which together with their
unusually long periods, distinguished them as a unique class of pulsars.
MLRH have now shown from an analysis of the SNRs associated with SGR/AXPs,
that essentially all ($\sim$ 80\%) of them occur in the Warm phase of the ISM.
This is very unexpected, since observations of extragalactic Type II/Ibc SN
(van Dyk, Hamuy \& Filippenko 1996), Galactic SNRs (Higdon \& Lingenfelter
1980) and young pulsars (MLRH), all show (Table I) that only a small
fraction ($<$ 20\%) of all the neutron star producing Type II/Ibc SN actually
occur in the Warm phase of the ISM, whereas the bulk ($>$ 80\%) of them occur
instead in the more tenuous ($n \sim$ 0.001 cm$^{-3}$) Hot phase. Consequently,
something about the dense ISM environment shapes the character of the SGR/AXPs
and makes them different from the more common neutron star population of radio
pulsars.

%ADD Figure 1 (i.e. Figure 3 from Marsden et al. 2001)
\begin{figure}
%\epsfysize=8cm % fix the y-dimension and scales x-dim. to y-dim.
%\epsfxsize=8cm % fix the x-dimension and scales y-dim. to x-dim.
% Feel free to do the choice you prefer but do not exceed the x-dimension
% of the text lines
\epsfxsize=\linewidth
%\epsfbox{rfig1_rot.ps}
\epsfbox{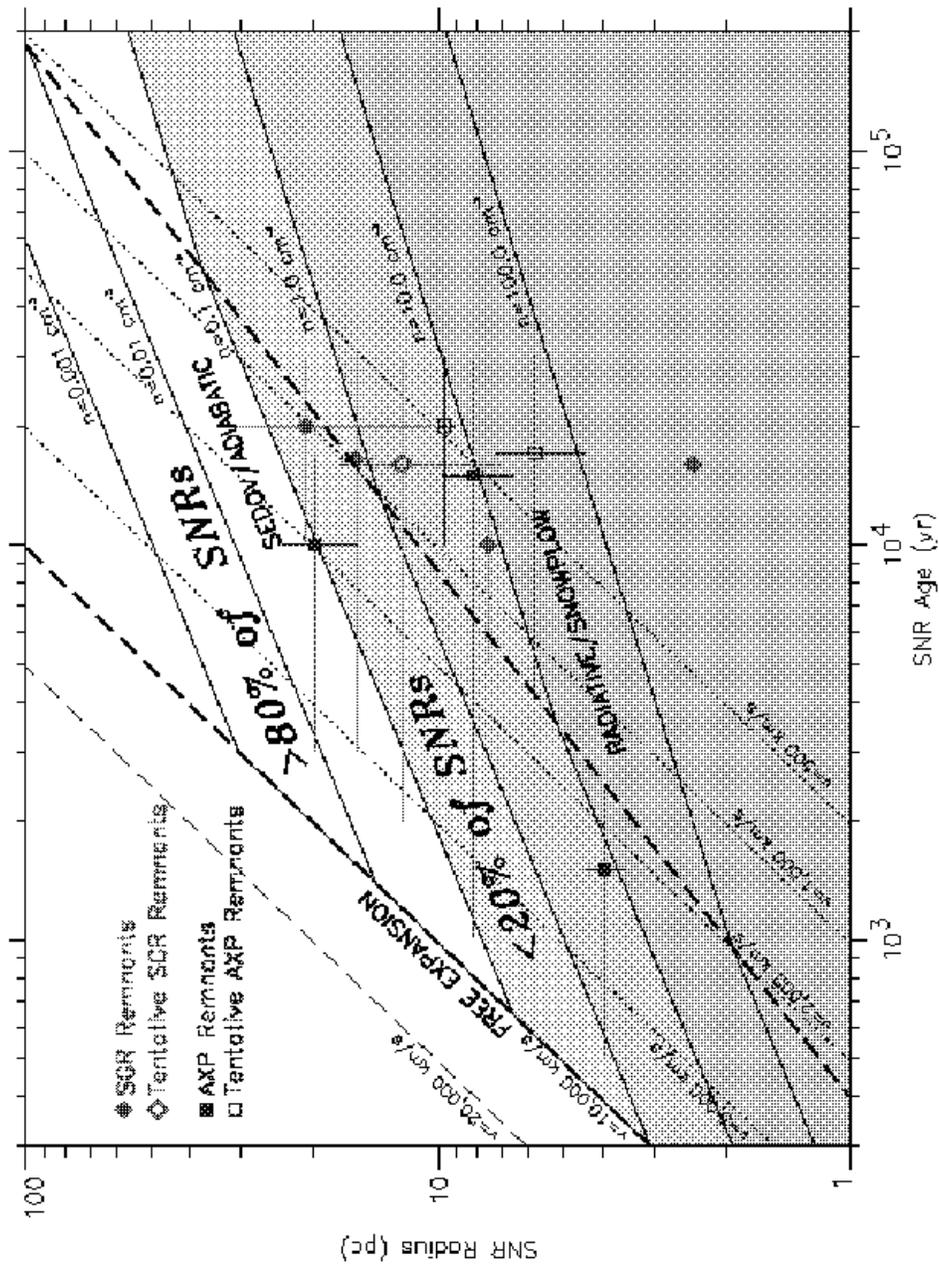}
\vspace*{-0.5cm}
\caption[h]{~The radius of the SGR and AXP supernova remnant shells as a
function of their age (from Marsden et al. 2001). The solid lines denote
SNR expansion trajectories in the free expansion, Sedov, and radiative
phases in a wide range of ISM densities. The dotted lines denote the tracks
of neutron stars born at the origin of the supernova explosion with varying
space velocities. The data show that these objects are unusual in that
they are all preferentially formed in the denser ($>0.1$ H cm$^{-3}$)
Warm phases of the interstellar medium (ISM), where $<$20\% of all neutron
star forming supernovae occur. As can be seen, the data are very robust even
though there are large uncertainties in the SNR ages.}
\end{figure}

%\vspace{1cm} %TO ALLOW SUFFICIENT SPACE BETWEEN THE TEXT AND THE FIGURES
\begin{center}
\begin{table}[h]
\caption{OCCURRENCES IN THE WARM AND HOT PHASES OF THE ISM}
\vspace*{.1in}
%\hspace{1.5cm} %if you want to center your table act on this argument
%\begin{tabular}{p{4cm}p{4cm}p{4cm}}
\hspace{2.5cm}\begin{tabular}{lcc}
\hline
\hline
\multicolumn{1}{c}{Source}& WISM(\%)$^a$ & HISM(\%)$^b$\\
\hline
Extragalactic~Supernovae$^c$ & $<$20 & $>$80\\
Galactic Supernovae$^d$      & 10$\pm$10 & 90$\pm$10\\
Young Pulsars$^e$            & 31$\pm$14 & 69$\pm$21\\
SGR/AXPs$^e$                 & 83$\pm$26 & 17$\pm$12\\
\hline
\end{tabular}\\
\hspace*{2.5cm}$^a$Percentage in the Warm ISM ($n \geq $ 0.1 cm$^{-3}$)\\
\hspace*{2.5cm}$^b$Percentage in the Hot ISM ($n \sim$ 0.001 cm$^{-3}$)\\
\hspace*{2.5cm}$^c$van Dyk, Hamuy \& Filippenko (1996) as discussed in\\
\hspace*{2.7cm}Marsden et al. (2001)\\
\hspace*{2.5cm}$^d$Higdon \& Lingenfelter (1980) as discussed in Marsden\\
\hspace*{2.7cm}et al. (2001)\\
\hspace*{2.5cm}$^e$Marsden et al. (2001)
\end{table}
\end{center}
\vspace{-1.0cm}

The unusually dense phases of the interstellar medium in which
the SGR/AXPs are born confine the progenitor winds and rapidly
slow the supernova ejecta, initiating a reverse shock, which
in turn reverses the flow of the innermost ejecta for capture
by the nascent neutron star. Inflow of this material provides an
additional torque along with magnetic dipole radiation (MDR) to
rapidly spindown the initially fast rotating neutron star to very
slow (several second) periods in $\sim 10^4$ years. Here we show
that the effects of such rapid spindown can also provide the energy
for the observed bursts through plate-tectonic driven crustal
subduction and phase transitions.

The unexpected occurrence of most, and perhaps all, of the SGR/AXPs
in the denser phases of the ISM effectively rules out the magnetar
model of their origin, since in that model supercritical magnetic
fields are an intrinsic property of the neutron star with no plausible
relation to the external environment. It has been suggested that
magnetars might form from progenitors with the largest angular momentum,
so that they might only be formed from the most massive stars, which,
because they evolve most rapidly, may explode preferentially in the
denser regions where they were formed. But this suggestion is not
tenable, since pulsar observations (e.g. Cordes \& Chernoff 1998)
show that there is no correlation between spin period and magnetic
field strength, and observations of giant star formation regions show
that only the most massive stars in the first of several generations
of star formation explode in the dense cloud environment while most
of those formed in later generations explode in the hot, low density
environment of the superbubble created by the earlier supernova
explosions (e.g. McKee \& Williams 1997).

\section{Propeller Spindown from Fossil Disks}

Propeller spindown models of the SGR/AXPs, on the other hand,
can be strongly influenced by the circumstellar environment.
The rapid spindown rates, young ages inferred from the SNR ages,
long spin periods clustered around 5-10 s, and $\sim$ 10$^{35}$ erg/s
x-ray luminosities can all be explained by models involving the
propeller effect on inflowing material (Illarionov \& Sunyaev 1975)
as the dominant spindown torque. Since no binary companions have been
detected around SGR/AXPs, the infalling material must come from fossil
disks which can spindown the neutron star on time scales of 1-10 kyr.

Fossil disks may be formed from the supernova ejecta being pushed back
toward the star by the reverse shock, which can actually reverse the
flow of the slowest moving inner ejecta (Truelove \& McKee 1999),
pushing it back toward the nascent neutron star to form a disk.
The occurrence of ``pushback" disks will depend on the strength of the
reverse shock, which forms in the Sedov phase of the SNR expansion from
the interaction between the supernova blast wave and the external gas
and is thus strongly affected by the density of the circumstellar ISM.
These disks are most likely to form around neutron stars born from the
more massive progenitors in the denser phases of the ISM which confine
the progenitor winds much nearer the star, so that the expanding SN
ejecta can sweep up gas and develop both forward and reverse shocks
much more rapidly. Such a situation can be seen in evolution of the
ejecta from SN 1987A, which is surrounded by very dense ($n \sim$
10$^2$ to 10$^3$) gas from confined progenitor winds (e.g. Chevalier
\& Dwarkadas 1995) and has already entered the Sedov phase dramatically
slowing the forward shock from 30,000 km/s to only 3,000 km/s within
10 yrs of the explosion. For such densities the pushback process should
begin at a ``reversal time" of 400 to 800 yr with an expected (Truelove
\& McKee 1999) pushback mass of $\sim$ 0.4 $M_{\odot}$ for a total ejecta
mass of 10 $M_{\odot}$. Only a very small fraction of the pushed back
ejecta is needed to form a fossil disk with the 10$^{-6} M_\odot$
required to explain the spindown of SGR/AXPs via the propeller mechanism.

The fossil disk will exert a spindown torque on the neutron star, if the
inflow rate is low and the magnetic field is strong, so that the majority
of the inflowing material is accelerated away in a bipolar wind which
carries off angular momentum from the magnetosphere, and hence from the
neutron star itself (e.g. Illarionov \& Sunyaev 1975).
For typical radio pulsar magnetic fields of $10^{12}$ Gauss, the
observed 6--12 s spin period range of SGR/AXPs is naturally explained
by mass infall rates at the magnetospheric boundary of 4--20$\times
10^{15}$ g/s. Such rates can also account for the observed x-ray
luminosities of $\sim 10^{35}$ ergs/s if only about 5-25\% of the
infalling material actually reaches the neutron star surface.

Propeller spindown from such disks can also easily account for the
observed number of SGR/AXPs. From the Galactic neutron star birth
rate (1/40 yr$^{-1}$), the fraction of neutron star progenitors
in the warm dense ISM ($<$0.2), and the fraction of such progenitors
($>$ 20 $M_{\odot}$) that suffer mass loss sufficient to form a pushback
disk ($0.1$), MLRH estimate that the number of SGR/AXPs formed in
the last 30 kyr should be $<$15, which is quite consistent with
the observed number of 12.

Propeller spindown can also explain many other quiescent aspects
of the SGR/AXPs in a natural manner. This includes spindown ages
comparable to the ages of the associated SNRs, the narrow range of
spin periods, the relatively low luminosities, and the lack of
prolonged periods of spin-up. Moreover, we suggest that the
exceedingly rapid spindown of these neutron stars provides both the
energy and mechanism for the very energetic bursts seen from SGRs.

\section{Mechanisms and Energetics of SGR Bursts from Propeller Spindown}

The similarities between the size-intensity distributions (Cheng et al.
1996) and other features (Palmer 1999) of SGR bursts and earthquakes
suggests that the physics of these two phenomena may be similar. With
this in mind, the magnetar model postulates that the majority of SGR
bursts are crust cracking events caused by the diffusion and possible
decay of the superstrong magnetar field. As exemplified by the Earth,
however, a superstrong magnetic field is not required to excite crustal
quakes, and one simple source of SGR bursts is quakes caused by neutron
star plate tectonics (Ruderman 1991), driven by the rapid spindown of
these objects. This process is illustrated in Figure 2. Rapid spindown
should cause segments of the neutron star's
crust to be cracked and dragged toward the rotational equator by pinned
superfluid vortices. Subduction of crustal material will occur as plates
collide and will release gravitational energy as the subducted crust
undergoes compressional phase transitions deep in the interior of the
star. The largest, deepest quakes on the Earth are caused by compressive
phase transitions of subducted crust. Vibrations excited by this process
will be transmitted into magnetospheric Alv\`{e}n waves which accelerate
particles --- producing x-ray/gamma-ray emission.

\begin{figure}
\epsfysize=6cm % fix the y-dimension and scales x-dim. to y-dim.
%\epsfxsize=8cm % fix the x-dimension and scales y-dim. to x-dim.
% Feel free to do the choice you prefer but do not exceed the
%x-dimension of the text lines
\centerline{\epsfbox{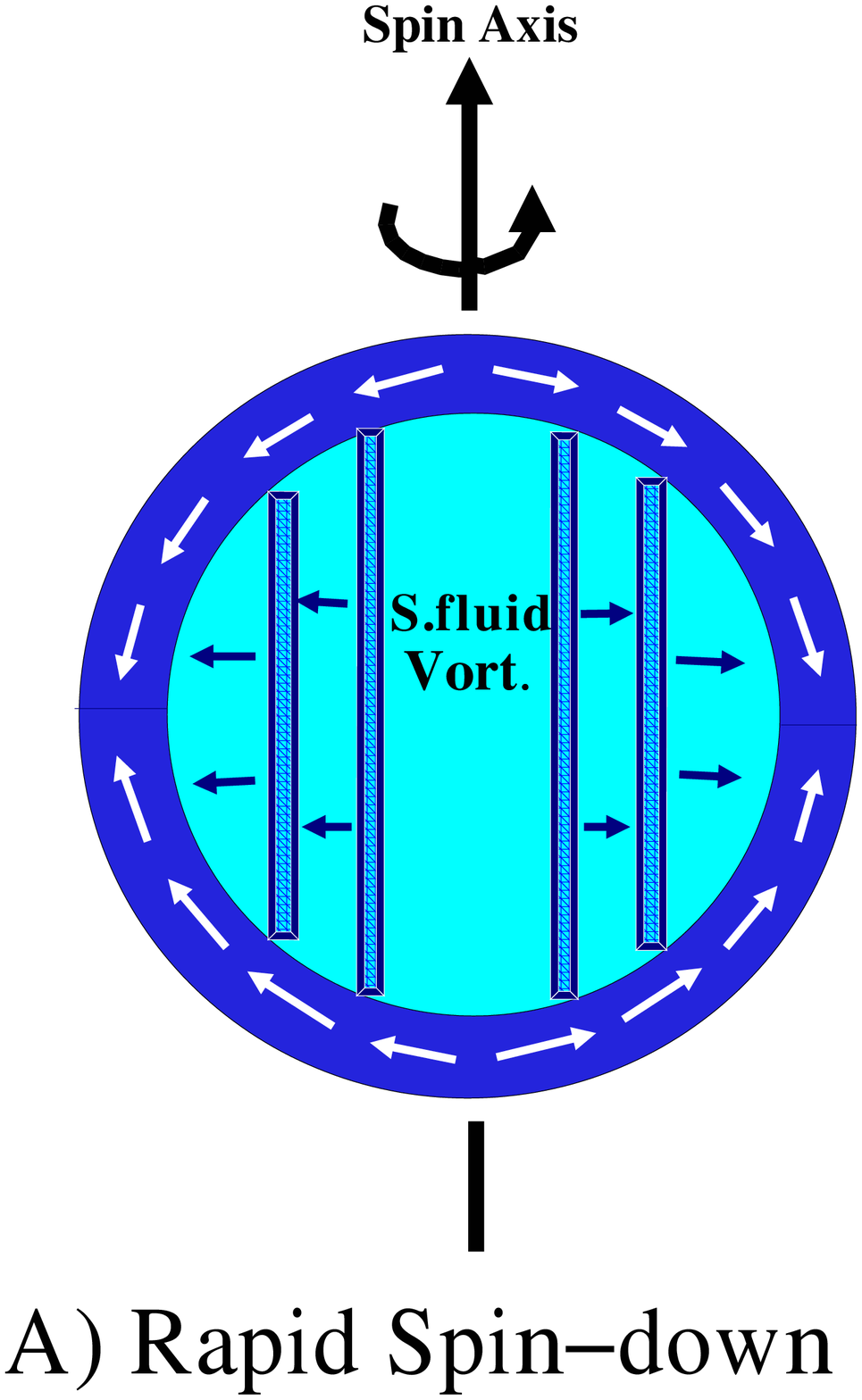}\epsfysize=6cm\epsfbox{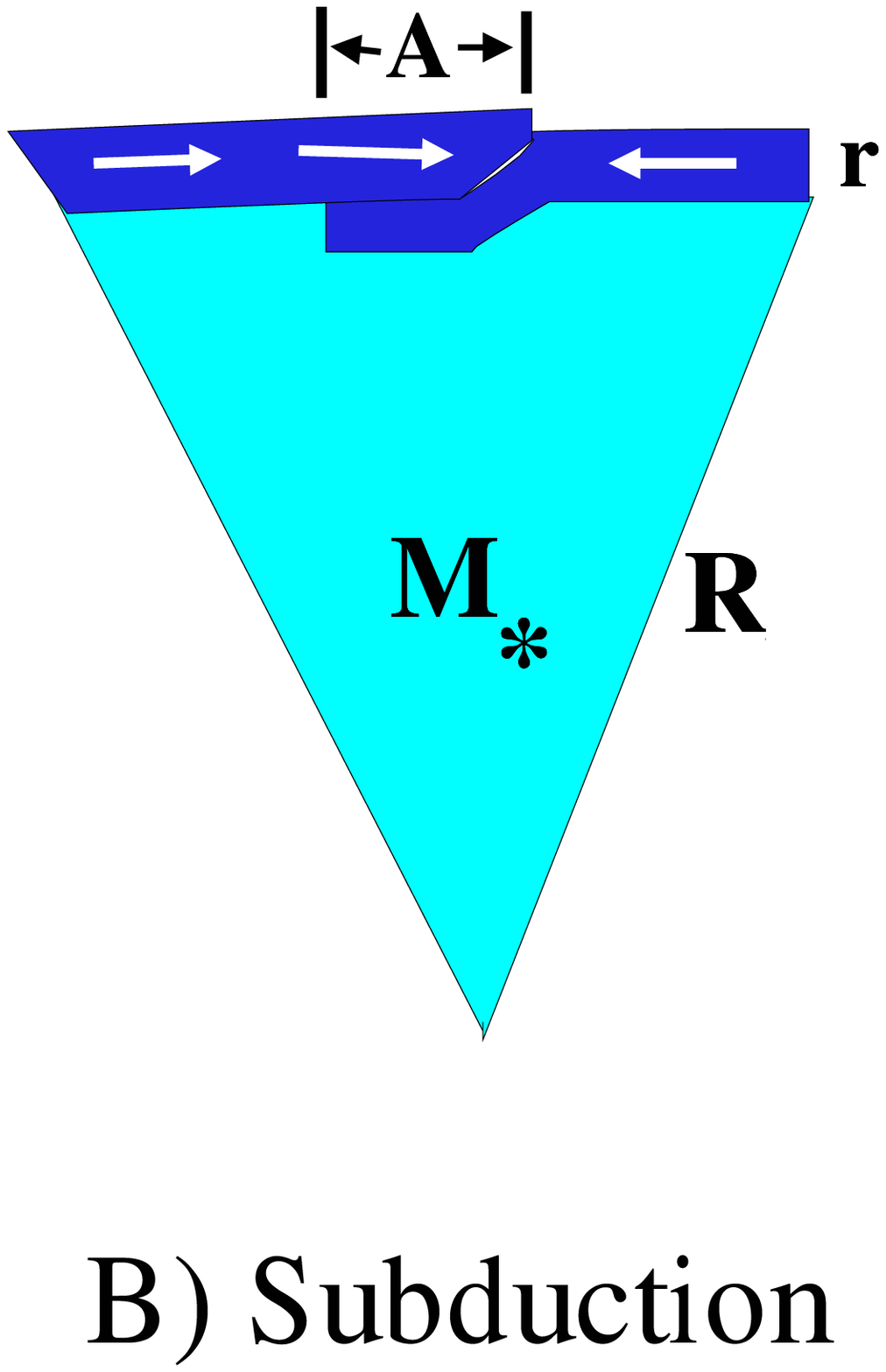}
\epsfysize=6cm\epsfbox{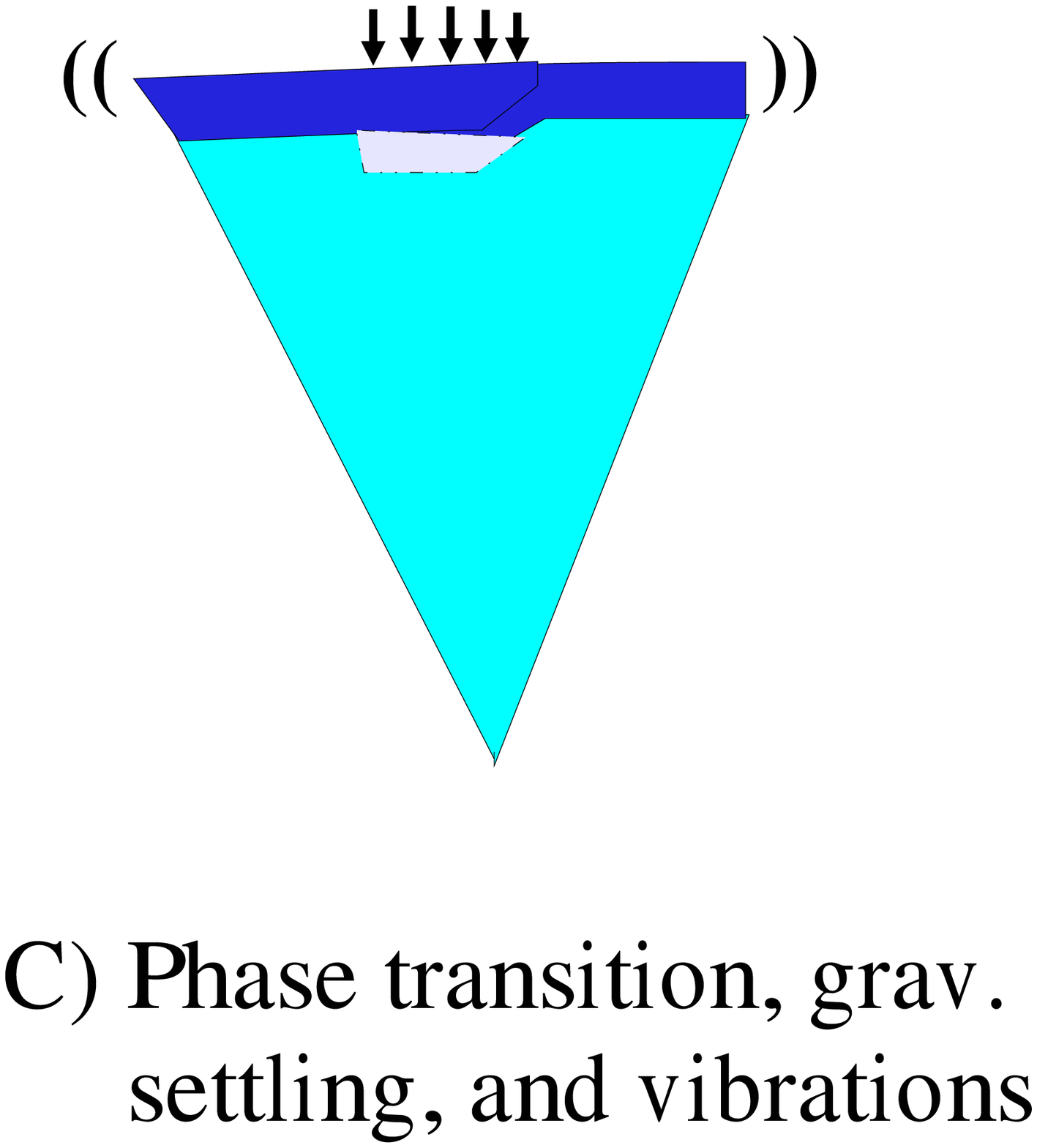}}
\caption[h]{~The phase transition starquake model for SGR bursts.
Panel A shows a slice along a plane parallel to the neutron star
spin axis, while Panels B and C show slices across the equatorial
plane of the star. See the text for details.}
\end{figure}

The energy released in the subduction process can be estimated and
compared with that of the most energetic SGR bursts. The subduction
of relatively light crust into the dense interior of the star will
eventually result in one or more phase transitions, when blocks of
low density crust are transformed into much denser interior phases.
The resulting decrease in volume of the subducted material will cause
settling of the overlying material, releasing gravitational energy
and exciting vibrations throughout the neutron star. The gravitational
energy released in this process by the settling of $\Delta M$ of
overlying material dropping by $\Delta R$ is $\Delta E_{g} \approx
(GM_{\ast}^2/R_{\ast})(\Delta M/M_{\ast})(\Delta R/R_{\ast})$, or
5$\times$10$^{53}(\Delta M/M_{\ast})(\Delta R/R_{\ast})$ ergs.
Thus for a crustal block of area equal to a fraction $A$ of the
surface of the star and thickness $r = \Delta R/R_{\ast}$ undergoing
a phase transformation, compressing it as little a 10\%, at a fractional
depth $d = D/R_{\ast}$, the overlying mass is $\Delta M \approx
A d M_{\ast}$ and $\Delta R \approx 0.1 r R_{\ast}$, so that the
gravitational energy released is 5$\times$10$^{43} A_{-3}d_{-3}r_{-3}$ ergs,
even for all of the fractions as small as 10$^{-3}$. This energy is easily
sufficient (e.g. Cheng et al. 1996) to power a single typical SGR burst
with an x--ray production efficiency of a few percent or less. The
superbursts, with isotropic energy releases of $\sim 10^{44}$ ergs
(e.g. Ramaty et al. 1980), require larger sections of the crust to
participate (or possibly a deeper core phase transition; e.g. Ellison
\& Kazanas 1983), but they are still explainable by this mechanism.
Thus the release of gravitational energy by spindown induced phase
transitions can explain the energetics of SGR bursts. The mean gravitational
energy density $(GM_{\ast}^2/R_{\ast})/V_{\ast}$ of $\sim$ 10$^{35}$
erg cm$^{-3}$ of subducted material in a neutron star greatly exceeds
the magnetar magnetic energy density in the same volume unless the
mean stellar magnetic field is unrealistically strong ($>$ 10$^{18}$
G). Thus gravitational energy is a much more plausible source of
SGR burst energy than magnetic energy.

Other characteristics of SGR bursts can similarly be explained in
terms of canonical neutron star parameters and conventional neutron
star physics. A superstrong magnetic field, for example, is {\it not}
required to explain the short durations of SGR bursts, which can be
explained quite easily in terms of the storage time for energy in
the neutron star crust (Blaes et al. 1989). The $\sim 100$ s duration
of the rare SGR superbursts can be plausibly attributed to the
gravitational radiation timescale from a vibrating neutron star
(Ramaty et al. 1980). The spectral hardness and luminosity of SGR
bursts can explained by a synchrotron cooling model (Ramaty, Bussard,
\& Lingenfelter 1981). Magnetar-strength magnetic fields are not
required in this model, as illustrated by the following argument
based on the March 5th 1979 superburst from SGR 0526--66.
This burst had a peak luminosity of $L_{p}\sim 5\times 10^{44}$
erg/s, but the synchrotron cooling time of the vibrationally-heated
electrons in a $\sim 10^{12}$ Gauss field is only $t_{s}\sim 2\times
10^{-16}\gamma_{5}^{-1}$ s, where we have assumed a relativistic
gamma factor of $\gamma=5\gamma_{5}$ for the emitting electrons
(e.g. Rybicki \& Lightman 1979). This means that only $L_{p}t_{s}
\sim 10^{29}$ ergs of electrons are emitting at any one time, which
can easily be confined into only a $1$ m cube by the $10^{12}$ Gauss field.
Finally, a thin fossil disk could survive the intense photon flux from
even the strongest SGR bursts, because the total energy deposited in
the disk from a burst would be less than the gravitational binding
energy of the disk for typical disk parameters.

\section{Conclusion}

We have shown that propeller spindown from fossil disks can explain
many of the features of SGRs and AXPs. The lack of binary companions,
rapid spindown, and SGR burst energetics and spectra can also be
explained by the magnetar model. But several very important properties
--- clustered spin periods, SNR ages, dense environments, and quiescent
energetics --- which are easily explained by the propeller disk model,
have not been explained in the context of the magnetar model.
These distinguishing characteristics are listed in Table II.

As mentioned previously, the clustering of SGR and AXP spin periods
in the range $5-12$ s is evidence of a characteristic spin period,
which is a natural consequence of propeller spindown from a fossil
disk (Chatterjee \& Hernquist 2000). There can be no spin equilibria
in magnetars, however, since magnetars can only spindown (Harding,
Contopoulis, \& Kazanas 1999). Therefore if AXPs and SGRs were magnetars,
the clustering of their spin periods would have to be a coincidence,
which is extremely unlikely. Similarly, the association of SGRs and AXPs
with middle age ($\sim 10$ kyr) SNRs is at odds with their $\sim 1$
kyr MDR timing ages if they are magnetars\footnote{The effects of
Alfv'{e}n wind torques (Harding, Contopoulis, \& Kazanas 1999) and
magnetic field decay (e.g. Colpi et al. 2000) in magnetars would
only increase this age discrepancy.}. Propeller spindown can explain
the MDR age discrepancies in SGRs and AXPs (Chatterjee, Hernquist,
\& Narayan 2000) and other sources (Marsden, Lingenfelter, \&
Rothschild 2001a,b), while providing a link between SGRs/AXPs and
other isolated neutron stars. As discussed above, the environments
of SGRs and AXPs also very strongly favor the propeller disk model,
while no plausible argument has been found as to how the development
of magnetars could be affected by their environment. Finally, the
broadband quiescent emission from SGRs (Kaplan et al. 2001) and AXPs
(Hulleman et al. 2000a) would require large scale magnetar fields of
$>10^{15}$ Gauss to supply the necessary energy. Such extremely large
magnetic fields would greatly exacerbate the magnetar age problem
discussed above, because of the extremely rapid spindown caused by
the magnetic torques. Assuming an x--ray efficiency of only $1\%$,
the propeller disk model requires a disk mass of $\sim 10^{-5}M_{\odot}$
to power the SGR and AXP quiescent emission for a lifetime of $10$ kyr.
This is a very small fraction of the ejecta mass pushed back by the
reverse shock in a massive Type II SN in the denser ISM.

%\vspace{1cm} %TO ALLOW SUFFICIENT SPACE BETWEEN THE TEXT AND THE FIGURES
\begin{center}
\begin{table}[h]
\caption{CONSISTENCY OF PROPELLER AND MAGNETAR MODELS WITH SGR/AXP PROPERTIES}
\vspace*{0.1in}
%\hspace{1.5cm} %if you want to center your table act on this argument
%\begin{tabular}{p{4cm}p{4cm}p{4cm}}
\hspace{1.5cm}\begin{tabular}{lcc}
\hline
\hline
SGR/AXP Property & Propeller Model & Magnetar Model\\
\hline
Clustered Spin Periods? & Yes & No\\
SNR Ages? & Yes & No\\
Dense Environments? & Yes & No\\
Quiescent Energetics? & Yes & No\\
\hline
\end{tabular}
%%%%%%%%%%%%%%%%%%%%%%%%%%%%%%%%%%%%%%%%%%%%%%%%%%%%%%%%%%%%%%%%%%%%%%%%%
\end{table}
%%%%%%%%%%%%%%%%%%%%%%%%%%%%%%%%%%%%%%%%%%%%%%%%%%%%%%%%%%%%%%%%%%%%%%%%%%
\end{center}

Although the propeller disk model is quite successful in explaining
the properties of SGRs and AXPs, there are two important issues that
need to be addressed. The first issue concerns the multiwavelength
emission from pushback disks, which should manifest itself primarily
at infrared wavelengths as result of the high fraction of refractory
dust grains from the heavy supernova ejecta. Because of the high
grain content, standard gas disk spectra (e.g. Perna, Hernquist,
\& Narayan 2000) are not applicable in this case, since the
bulk of the disk energy will be radiated at wavelengths longer
than the optical, where restrictive upper limits have been obtained
for some of these objects (Hulleman et al. 2000b; Kaplan et al. 2001).
Sensitive IR observations of the nearby SGRs and AXPs are needed
before serious constraints can be placed on the propeller disk model.
Secondly, the recently noted correlation between the spectra and
spindown torques of SGRs and AXPs (Marsden \& White 20001) needs
to be explained in terms of the two theoretical models for SGRs and
AXPs. Models of accreting neutron star spectra can produce spectra
with the blackbody plus soft power law shape characteristic of SGRs
and AXPs, but it is unclear if the spectral hardness increases with
spindown torque as indicated by the observations. Similarly, detailed
calculations of the quiescent spectra of magnetars have yet to be done.

\acknowledgements
This work was performed while one of the authors (DM) held a
National Research Council-GSFC Research Associateship. RER
acknowledges support by NASA contract NAS5-30720, and REL support
from the Astrophysical Theory Program.

% References. We avoided using the \bibitem commmand since we found it is
% somewhat platform-dependent. We also avoided using the \cite{keyword}
% command since we found it cumbersome. However, if you are an expert
% LateX user you may use the various LateX tools for the references
% provided they give the same printout formats of the examples given here.

\end{document}